\documentclass[11pt]{article}

\usepackage[preprint]{acl}

\usepackage{times}
\usepackage{latexsym}

\usepackage[T1]{fontenc}

\usepackage[utf8]{inputenc}

\usepackage{microtype}

\usepackage{inconsolata}

\usepackage{graphicx}
\usepackage{amsmath}
\usepackage{booktabs}
\usepackage{multirow}
\usepackage{pifont}
\newcommand{\cmark}{\ding{51}} 
\newcommand{\xmark}{\ding{55}} 
\newcommand{\pmark}{\ding{108}} 

\title{AuditFraudBench: Benchmarking Audit Judgment in \\Detecting Fraudulent Misstatements}

\author{%
  Zhiwei Liu\textsuperscript{1}\quad
  Yueru He\textsuperscript{2}\quad
  Qing Ou\textsuperscript{3}\quad
  Tianlei Zhu\textsuperscript{2}\quad 
  \textbf{Xiaorui Guo}\textsuperscript{4}\quad \\
  \textbf{Xueqing Peng}\textsuperscript{5}{\thanks{Corresponding Author}}\quad 
  \textbf{Sophia Ananiadou}\textsuperscript{\textbf{1}} \\ 
    \textsuperscript{1}The University of Manchester \quad 
    \textsuperscript{2}Columbia University \quad 
    \textsuperscript{3}Rutgers University \quad \\
    \textsuperscript{4}The University of Edinburgh \quad
    \textsuperscript{5}The Fin AI \quad \\
\texttt{\{zhiwei.liu,sophia.ananiadou\}@manchester.ac.uk} \\
\texttt{xueqing.peng2024@gmail.com}
}

\begin{document}
\maketitle
\begin{abstract}
Large language models (LLMs) have shown strong performance in financial analysis and surface-level factual error detection, yet their ability to identify fraudulent financial misinformation in audited corporate reporting remains underexplored. Existing financial and audit benchmarks mainly focus on factual verification, numerical reasoning, rule compliance, or audit workflows, but rarely evaluate misleading disclosure narratives or management explanations that obscure the true drivers of reported performance. We introduce AuditFraudBench, an enforcement-grounded benchmark constructed from authentic company filings and regulatory materials, including original and restated 10-K and 10-Q filings, structured financial statements, MD\&A disclosures, and SEC Accounting and Auditing Enforcement Releases (AAERs). AuditFraudBench contains three tasks: Profit Source Attribution, Misleading Narrative Detection, and Fraud Pattern Classification, which evaluate whether models can identify the true source of reported performance, detect misleading disclosure framing, and classify misconduct mechanisms into known manipulation patterns. We evaluate GPT, DeepSeek, and Qwen series LLMs on the benchmark. Results show that both proprietary and open models still struggle to jointly reason over financial figures, disclosure framing, restatement evidence, and enforcement-grounded fraud mechanisms. AuditFraudBench provides a challenging testbed for audit-relevant, evidence-grounded evaluation of LLMs in financial reporting.

\end{abstract}

\section{Introduction}

Although large language models (LLMs) have shown strong performance in financial analysis \cite{nie2024survey}, forecasting \cite{NEURIPS2024_adb1d9fa,peng2025multifinbenbenchmarkinglargelanguage}, and the detection of surface-level factual or numerical errors \cite{peng2026herculeanagenticbenchmarkfinancial}, their ability to detect fraudulent misstatements remains underexplored. Unlike simple factual, numerical, or rule-compliance errors, fraudulent misstatements consist of plausible and seemingly compliant disclosures that obscure distorted performance drivers, opportunistic accounting choices, or omitted risk-relevant context, raising the question of whether LLMs can detect disclosures strategically framed to mislead. This problem is especially important in financial reporting, where misleading disclosures can distort investor decisions and conceal managerial misconduct \cite{celestin2015uncovering}. Fraudulent misstatements are often intentionally disguised as plausible and compliant reporting: figures may be internally consistent, while management attributes performance changes to sustainable operations when the true driver is revenue timing, expense deferral, reserve release, one-time gains, or accounting estimate manipulation \cite{reurink2019financial,schilit2010financial}. Likewise, MD\&A narratives may mislead not through explicit falsehoods, but through selective emphasis, omitted context, or favorable framing \cite{1151a90b345f43d287242021d0e904c5}. Detecting such misinformation, therefore, requires models to connect financial figures, managerial explanations, accounting choices, and known fraud mechanisms, a form of reasoning closely aligned with audit judgment and professional skepticism. As LLMs are increasingly used to support financial analysis, audit assistance, investment research, and corporate disclosure review, systematically evaluating whether they can recognize such high-risk, intentionally disguised misinformation has become both timely and necessary.

Existing benchmarks cover related but incomplete aspects of this challenge, focusing mainly on fact verification, numerical consistency, and rule-compliance checking rather than the harder task of detecting fraudulent misstatements and strategically misleading disclosures. Audit-oriented benchmarks, including FinAuditing \cite{wang2025finauditing}, Automating Financial Statement Audits \cite{wang2025automating}, FinRule-Bench \cite{vignesh2026finrule}, and FinMaster \cite{jiang2025finmaster}, assess LLMs' ability to process filings, reason over accounting structures, detect rule violations, and complete calculation- or workflow-based audit tasks. \textit{However, they do not directly evaluate whether models can detect misleading disclosure narratives or distinguish management's stated explanations from the true drivers of reported performance.} Financial misinformation benchmarks, such as MFMD-Scen \cite{liu2026same}, FinFact \cite{rangapur2025fin}, FinDVer \cite{zhao2024findver}, FISCAL \cite{sharma2025fiscal}, and RFCBench \cite{jiang2026all}, examine factual reliability, claim verification, and inconsistency detection in financial contexts, but are generally not grounded in audited filings or enforcement cases. As a result, they do not capture the distinctive mechanisms of corporate disclosure manipulation, including revenue timing, expense deferral, earnings smoothing, accounting estimate manipulation, and misleading MD\&A framing. \textit{This leaves a gap in evaluating whether LLMs can identify audit-relevant financial misinformation that is financially grounded yet rhetorically disguised.} A more detailed discussion of related work is provided in Appendix \ref{sec:relatedwork}.

To address this gap, we propose AuditFraudBench, an enforcement-grounded benchmark for evaluating LLMs' ability to detect fraudulent financial misinformation in audited corporate reporting. AuditFraudBench is constructed from authentic company filings and regulatory enforcement materials, including original 10-K and 10-Q reports, restated filings, structured financial statements, MD\&A disclosures, and SEC Accounting and Auditing Enforcement Releases (AAERs). Rather than treating financial misinformation as generic factual inconsistency, AuditFraudBench evaluates three complementary capabilities required for audit-relevant fraud detection: identifying whether reported performance is attributed to the correct underlying driver, detecting misleading disclosure narratives that obscure material context, and recognizing the fraud mechanism documented in enforcement actions. These tasks jointly require models to connect reported and restated financial figures, managerial explanations, disclosure framing, and AAER-grounded misconduct patterns, thereby moving beyond fact verification, numerical checking, and rule-compliance benchmarks. We evaluate GPT, DeepSeek, and Qwen series LLMs on AuditFraudBench. Experimental results show that the benchmark poses substantial challenges for current LLMs, as strong proprietary and open models alike exhibit limited ability to jointly reason over financial evidence, narrative framing, and enforcement-grounded fraud mechanisms.

Our main contributions are as follows:
(1) We introduce AuditFraudBench, an enforcement-grounded benchmark for evaluating LLMs' ability to detect fraudulent financial misinformation in audited corporate reporting, with a focus on misleading narratives and disguised accounting manipulation.
(2) We design three complementary tasks: Profit Source Attribution, Misleading Narrative Detection, and Fraud Pattern Classification, which respectively evaluate whether models can identify the true source of reported performance, recognize misleading MD\&A framing, and map misconduct mechanisms to known fraud patterns.
(3) We construct the benchmark from authentic financial reporting and enforcement sources, including original and restated filings, structured financial statements, MD\&A passages, and SEC AAERs, enabling evidence-grounded evaluation that better reflects audit judgment and investor-relevant disclosure analysis.

\section{AuditFraudBench}

\begin{figure*}[htb]
\centering
  \includegraphics[width=1.7\columnwidth]{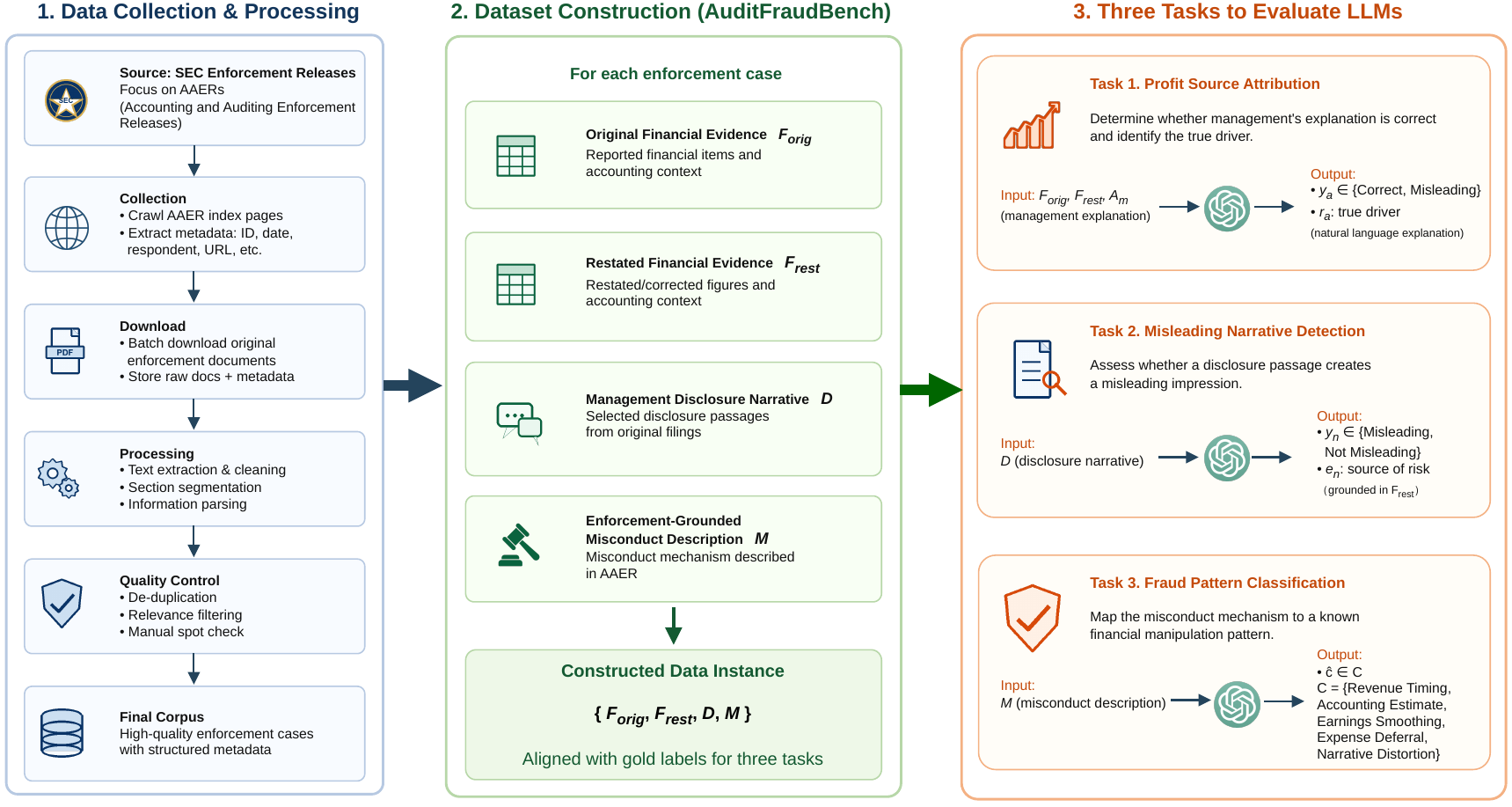}
  \caption{ Overview of the AuditFraudBench construction.}
  \label{fig:mainmethod}
\end{figure*}

Figure \ref{fig:mainmethod} present the construction of AuditFraduBench. In this section, we first present the task formulation in Sec \ref{sec:taskformulation}. We then describe the data collection and benchmark construction pipeline, including raw data acquisition, preprocessing, quality control, and data partitioning (Sec \ref{sec:datacollection}), followed by the detailed design of each benchmark task (Sec \ref{sec:taskdesign}).

\subsection{Task Formulation \label{sec:taskformulation}}

Within the context of LLMs, each data item in AuditFraudBench is constructed from enforcement-grounded financial reporting evidence. Formally, each instance consists of the original filing evidence $F_{orig}$, the restated or corrected financial evidence $F_{rest}$, the management disclosure narrative $D$, and the enforcement-grounded misconduct description $M$. The original and restated financial evidence contain a set of financial items $\{x_1, x_2, ..., x_n\}$, where each item may include a reported value, a restated value, and its corresponding accounting context. The disclosure narrative $D$ consists of $k$ sentences $\{s_1, s_2, ..., s_k\}$. Based on these inputs, AuditFraudBench evaluates three complementary tasks.

\textbf{Task 1. Profit Source Attribution.} The aim is to determine whether management's stated explanation for a reported performance change is correct, and to identify the true source of the change when the explanation is misleading. Given the original financial evidence $F_{orig}$, restated evidence $F_{rest}$, and management's stated explanation $A_m$, the model predicts an attribution decision $y_a$ and a true driver $r_a$.

\vspace{-5mm}

\begin{equation}
\small
(y_a, r_a) = \arg\max_{y,r} P_{\mathrm{LLM}}(y,r \mid F_{orig}, F_{rest}, A_m),
\end{equation}
where $y_a \in \{\text{Correct}, \text{Misleading}\}$ indicates whether the management attribution is valid, and $r_a$ is a natural language explanation of the true driver of the reported performance change, grounded in the accounting manipulation revealed by the restatement and Accounting and Auditing Enforcement Releases. 

\textbf{Task 2. Misleading Narrative Detection.} The aim is to assess whether an original disclosure passage creates a misleading impression, even if the passage is not explicitly false in isolation. Given the disclosure narrative $D$, the model predicts a misleading-risk label $y_n$ and an explanation $e_n$.

\begin{equation}
\small
(y_n, e_n) = \arg\max_{y,e} P_{\mathrm{LLM}}(y,e \mid D),
\end{equation}
where $y_n \in \{\text{Misleading}, \text{Not Misleading}\}$ indicates whether the disclosure creates a misleading risk
, and the explanation $e_n$ identifies the source of the misleading risk, such as selective emphasis, omitted context, favorable framing, or inconsistency with restated figures. $e_n$ is grounded in the restated filing evidence 
$F_{rest}$, specifically the restatement disclosure that reveals what the original passage omitted or misrepresented.

\textbf{Task 3. Fraud Pattern Classification.} The aim is to map a concrete misconduct mechanism to a known category of financial manipulation. Given the enforcement-grounded misconduct description $M$, the model predicts a fraud pattern category $c$ from a predefined label set $\mathcal{C}$.

\vspace{-5mm}

\begin{equation}
\footnotesize
\begin{split}
\mathcal{C} = \{\text{Revenue Timing},\ \text{Accounting Estimate},\ \text{Earnings} \\ \text{Smoothing},\  \text{Expense Deferral},\  \text{Narrative Distortion}\}.
\end{split}
\end{equation}

\vspace{-2mm}

\begin{equation}
\small
\hat{c} = \arg\max_{c \in \mathcal{C}} P_{\mathrm{LLM}}(c \mid M).
\end{equation}

Together, these three tasks evaluate whether LLMs can reason across the numerical, narrative, and mechanism levels of financial misinformation. Profit Source Attribution asks why the reported performance changed, Misleading Narrative Detection asks whether the company's disclosure creates a misleading impression, and Fraud Pattern Classification asks what type of manipulation pattern the misconduct represents.

\subsection{Data Collection \label{sec:datacollection}}

\subsubsection{Raw data}

This study takes enforcement disclosure texts issued by the U.S. Securities and Exchange Commission (SEC)\footnote{https://www.sec.gov/} as its core source corpus, with a primary focus on constructing supervised samples from Accounting and Auditing Enforcement Releases (AAERs)\footnote{https://www.sec.gov/enforcement-litigation/accounting-auditing-enforcement-releases}. AAERs are selected because their texts typically contain the investigated entity, the period during which the violation occurred, the accounting treatment involved, the financial impact of the misconduct, and the final enforcement conclusions. As such, they provide a natural comparative framework for examining whether managerial narratives are consistent with underlying financial facts. In other words, AAERs, as ex post regulatory determinations, serve as highly credible supervisory signals for tasks concerning the authenticity of financial narratives.

\subsubsection{Collection Process}

\paragraph{Data Collection.} During the data collection stage, the study first crawls case entries from the SEC AAER official index pages and extracts case-level metadata, including AAER number, release date, respondent, release identifier, and document URL. The original enforcement documents are then downloaded in batches, and corresponding metadata sidecar files are stored for each document, forming a dual-track storage structure consisting of “raw documents + structured metadata.” To ensure collection stability, the crawling process incorporates fixed request intervals, exception retries, rate-limit backoff, and periodic checkpoint writing mechanisms, thereby reducing the risk of sample loss caused by network fluctuations or platform access restrictions. The output of this stage is a traceable raw enforcement corpus rather than a modeling-ready dataset.

\paragraph{Cleaning.} After ingestion, the raw corpus undergoes document quality control and cleaning procedures. These include detecting corrupted files, parsing failures, and abnormally short texts (e.g., scanned documents without extractable text), while problematic samples are separately logged. When necessary, unusable samples are removed from downstream processing. The purpose of this step is not to maximize sample size, but rather to ensure that subsequent extraction and annotation are based on readable and machine-parsable texts, thereby reducing noisy labels and erroneous matches at the source.

\paragraph{Case Identification.} During the text processing and case identification stage, the study adopts a layered strategy combining “rule-based features + semantic understanding.” First, domain-specific keywords and pattern expressions are used to identify potential accounting distortion signals, such as revenue recognition issues, expense deferrals, accounting estimates, earnings smoothing, and misleading narratives. Second, structured text segmentation (e.g., summaries, findings of fact, and conclusion sections) is utilized to improve the contextual quality of semantic judgment. Finally, Deepseek is introduced to perform case-level semantic classification and field extraction. Compared with approaches relying solely on keywords, this strategy more effectively distinguishes substantive accounting manipulation cases from procedural or non-substantive enforcement matters, thereby improving the task relevance of candidate samples.

\paragraph{Structured Enhancement.} Based on the identification results above, the study further performs structured information enhancement via a dedicated LLM-based extraction pipeline to extract and normalize downstream-required fields, including mappable issuer entities, affected fiscal periods, descriptions of specific manipulation mechanisms, monetary impact, sanctioned party type (issuer/auditor/mixed), and assessments of whether the misconduct can be traced back to public financial statements. The key function of this step is to transform regulatory enforcement narratives into cross-source alignable data representations, enabling AAER texts to be automatically matched with EDGAR disclosure documents at the levels of “entity–period–document type.”

\paragraph{Sample Filtering.} During the candidate sample filtering stage, the study applies a set of hard constraints, retaining only cases that satisfy the following conditions: acceptable text quality, substantive new enforcement actions, relevance to the research task, mappability to a public issuer, clearly specified periods, sufficiently detailed manipulation mechanisms traceable to financial reports, and minimum confidence thresholds. After hard filtering, samples are ranked according to confidence scores, materiality of financial impact, completeness of period information, and related indicators to form a high-quality sample pool. In essence, this process further narrows the set of “readable enforcement texts” into “valid cases suitable for supervised comparative construction.”

\paragraph{Alignment and Pairing.} During the cross-source alignment and document pairing stage, selected cases are mapped to the SEC EDGAR system, where
issuer identifiers are resolved and original filings (Form 10-K) and, where available, corresponding amended filings (Form 10-K/A) for the relevant fiscal periods are retrieved. Where a restatement filing is located on EDGAR, the case yields a complete "original filing--restated filing" pair supporting all three benchmark tasks. Where only the original filing is found, the record still contributes to Tasks 1 and 3, while the Task 2 is populated with a structured placeholder. This pairing mechanism is central to the task design of the study because it places numerical and narrative disclosures from different filing versions of the same entity and period into a unified comparative framework, thereby supporting systematic modeling of the differences between “contemporaneous managerial explanations” and “subsequent corrective disclosures.”

\paragraph{Sample Construction.} During the sample construction stage, the study integrates two complementary categories of supervisory signals. The first category consists of structured financial fact signals: key indicators such as revenue, receivables, net income, and gross profit are extracted from XBRL company facts and aligned by reporting period and filing version to produce comparable numerical differences between original and restated filings. Because SEC XBRL reporting requirements were phased in progressively after 2009, structured financial figures are unavailable for cases whose relevant fiscal periods predate the XBRL mandate; such records contribute to narrative and classification tasks but lack structured numerical comparison. The second category consists of narrative evidence signals: Management’s Discussion and Analysis (MD\&A) sections are extracted from original filings, corrective explanations are extracted from amended filings, and AAER mechanism information is incorporated to generate task-oriented annotations. The final dataset is organized into three task categories:

Profit source attribution identification (whether managerial attribution is consistent with actual drivers);
Misleading narrative detection (locating contradictions between original narratives and corrective disclosures);
Accounting manipulation pattern classification (mapping cases to standardized fraud mechanism categories).


To improve the usability of the final dataset, the study introduces quality flags and targeted repair mechanisms during the post-processing stage. Samples containing placeholder text, suspected extraction errors, missing evidence, or inconsistent sources are automatically flagged and compiled into manual review lists. For repairable samples, only the relevant extraction modules are rerun instead of rebuilding the entire pipeline, thereby balancing quality and efficiency. Through a dual-layer control framework combining “automated rule-based quality inspection + manual spot checking,” the study ultimately constructs a reproducible, auditable, and extensible benchmark dataset.

\subsection{Task Design \label{sec:taskdesign}}


AuditFraudBench is constructed from enforcement-grounded financial reporting cases. Each case is derived from authentic company disclosures and regulatory evidence, including original 10-K filings, restated financial statements (Form 10-K/A), MD\&A passages, and SEC Accounting and Auditing Enforcement Releases (AAERs). For each case, we extract the financial figures, management explanations, disclosure narratives, and misconduct descriptions needed to support the three annotation tasks. 

\subsubsection{Task 1: Profit Source Attribution}

For each case, annotators examined the reported and restated financial evidence together with management's stated explanation for changes in earnings or operating performance. The goal was to determine whether the management explanation correctly attributed the reported performance change, or whether it obscured the true source of the change.

Annotators first compared the original reported figures with the restated or corrected figures, focusing on items that materially affected income, revenue, expenses, reserves, or other performance-related measures. They then assessed whether management's stated explanation, such as improved demand, higher operating efficiency, or business growth, was supported by the corrected financial evidence and enforcement record. If the explanation was supported, the instance was labelled \textit{Correct}. If the explanation attributed performance to an incomplete, unsustainable, or inaccurate source, the instance was labelled \textit{Misleading}, and annotators identified the true driver of the performance change.

The true driver was recorded as a concise accounting or economic cause, such as premature revenue recognition, expense deferral, reserve release, one-time gain, or accounting estimate manipulation. This task is designed as an attribution reasoning task rather than a numerical extraction task: financial figures provide the evidence, but the central objective is to judge whether the stated explanation faithfully represents the underlying source of reported performance.

\subsubsection{Task 2: Misleading Narrative Detection}

For each case, annotators reviewed original disclosure passages, typically drawn from MD\&A sections or related management discussion in 10-K and 10-Q filings. The goal was to determine whether the passage created a misleading impression for investors, even when the text was not explicitly false in isolation.

Annotators evaluated the disclosure passage in light of the relevant enforcement-grounded evidence. A passage was labelled \textit{Misleading} if it selectively emphasized favorable explanations, omitted material context, downplayed unsustainable or non-recurring drivers, or framed performance in a way that obscured the underlying misconduct. For example, a disclosure could be misleading if it attributed revenue growth to customer demand while omitting that the growth depended on premature recognition, extended payment terms, channel stuffing, or other non-normal sales practices. A passage was labelled \textit{Misleading} if it creates such a distorted impression.

For each \textit{Misleading} instance, annotators provided a brief explanation identifying the source of the misleading risk, such as selective disclosure, omitted context, favorable framing, or inconsistency with the enforcement-grounded misconduct mechanism. The gold explanation was grounded in the corresponding AAER or restatement evidence. This task captures narrative misinformation in corporate reporting, where misleadingness often arises from what the company emphasizes or omits rather than from a directly false statement.

\subsubsection{Task 3: Fraud Pattern Classification}

For each case, annotators classified the enforcement-grounded misconduct mechanism into a predefined financial manipulation category. The input to this task was the specific misconduct description extracted from AAERs, restatement disclosures, or related enforcement materials. We define the following possible labels:

\textbf{(1) Revenue Timing}: Assigned when the misconduct primarily involves improper timing or recognition of revenue, such as premature revenue recognition, fictitious sales, bill-and-hold transactions, channel stuffing, or recognizing revenue before required criteria were met.
\textbf{(2) Accounting Estimate}: Assigned when the misconduct primarily involves manipulation of accounting estimates or judgments, such as improper reserve releases, allowance manipulation, depreciation or amortization estimate changes, impairment judgments, or other discretionary estimates used to affect reported earnings.
\textbf{(3) Earnings Smoothing}: Assigned when the misconduct primarily involves shifting income or expenses across periods to reduce earnings volatility or meet performance targets, including cookie-jar reserves or other practices designed to smooth reported results.
\textbf{(4) Expense Deferral}: Assigned when the misconduct primarily involves delaying, capitalizing, or otherwise understating expenses or losses, such as improper capitalization of costs, failure to record liabilities, delayed impairment recognition, or deferral of ordinary operating expenses.
\textbf{(5) Narrative Distortion}: Assigned when the misconduct primarily involves misleading disclosures, omission of material information, or deceptive narrative framing intended to distort investors’ understanding of the firm’s performance, risks, or financial condition.


For cases involving multiple forms of misconduct, annotators selected the dominant mechanism based on the enforcement record and the primary impact on reported performance. This task evaluates whether models can move beyond case-specific textual understanding and recognize recurring financial manipulation patterns that generalize across enforcement cases.

Table \ref{tab:statistics} summarizes the statistics of AuditFraudBench. Since the dataset is derived from SEC filings and AAERs, all Task 1 instances involve attribution errors, and all Task 2 narratives are misleading.


\begin{table}[]
\footnotesize
\resizebox{0.48\textwidth}{!}{
\begin{tabular}{lclc}
\hline
Item              & N   & Item                 & N                    \\ \hline
Companies         & 84  & REVENUE TIMING       & 110                  \\
Unique AAER cases & 84  & ACCOUNTING ESTIMATE  & 107                  \\
Fiscal period     & 85  & EXPENSE DEFERRAL     & 37                   \\
10-K filings      & 143 & NARRATIVE DISTORTION & 23                   \\
10-Q filings      & 136 & EARNINGS SMOOTHING   & 18                   \\
20-F              & 16  & Total                & 295                  \\ \hline
Task 1            & 34  & Task 2               & 294                  \\
Task 3            & 295 &                      & \multicolumn{1}{l}{} \\ \hline
\end{tabular}
}
\caption{Data statistics. ``N'': number.\label{tab:statistics}}
\end{table}

\vspace{-2mm}



\section{Evaluations}

\subsection{Models}


Our evaluation of open-source and proprietary LLMs used the following reasoning-focused models: GPT-5.5, DeepSeek-V4-Pro, and Qwen3.5 reasoning variants (8B-R and 32B-R). We also assessed a diverse set of non-reasoning LLMs, i.e., GPT-4.1 \cite{openai_GPT41}, DeepSeek-V4-flash, the Qwen3.5 non-reasoning models (8B and 32B). The templates for evaluating LLMs can be found in Appendix \ref{sec:prompttemplates}. The open-source LLMs are evaluated on two NVIDIA Tesla L40S GPUs with 48 GB of memory. The temperature is set to 0, while all other settings use the default
configuration.

\subsection{Evaluation Metrics}

We use metrics such as Accuracy, Precision, Recall, and Macro-F1 for misinformation detection (classification) evaluation and ROUGE (1 and L) \cite{lin2004rouge} for explanation evaluation. We report ROUGE scores conditioned on the correctness of the predicted label. Specifically, ROUGE-conditional is computed only over examples where the predicted label matches the gold label, measuring explanation quality when the classification decision is correct. In contrast, ROUGE-gated assigns a ROUGE score of 0 to examples with incorrectly predicted labels and averages over all examples, providing an end-to-end measure of joint label prediction and explanation generation. This prevents explanations attached to incorrect labels from receiving inflated scores due to superficial lexical overlap with the reference explanation.

\section{Evaluation Results}

\begin{table*}[]
\resizebox{\textwidth}{!}{
\begin{tabular}{lcccccclcccccclcccccc}
\hline
\textbf{Model}             & \multicolumn{6}{c}{\textbf{Task 1: Profit Source Attribution}}                                                        &  & \multicolumn{6}{c}{\textbf{Task 2: Misleading Narrative Detection}}                                                                          &  & \multicolumn{6}{c}{\textbf{Task 3: Fraud Pattern Classification}}                                                                          \\ \cline{2-7} \cline{9-14} \cline{16-21} 
                  & Acc   & F1    & r1-c       & rL-c       & r1-g       & rL-g       &  & Acc            & F1             & r1-c       & rL-c       & r1-g       & rL-g       &  & Acc            & F1             & r1-c       & rL-c       & r1-g       & rL-g       \\ \hline
Deepseek-v4-flash & 1.000 & 1.000 & 0.235          & 0.171          & 0.235          & 0.171          &  & 0.065          & 0.061          & 0.285          & 0.205          & 0.018          & 0.013          &  & 0.420          & 0.355          & \textbf{0.138} & 0.114          & 0.058          & 0.048          \\
Deepseek-v4-pro   & 0.971 & 0.493 & 0.209          & 0.148          & 0.203          & 0.144          &  & 0.058          & 0.055          & 0.254          & 0.192          & 0.015          & 0.011          &  & 0.495          & 0.319          & 0.129          & 0.112          & 0.064          & 0.055          \\
GPT-5.5           & 1.000 & 1.000 & \textbf{0.249} & \textbf{0.172} & \textbf{0.249} & \textbf{0.172} &  & 0.041          & 0.039          & 0.281          & 0.195          & 0.011          & 0.008          &  & \textbf{0.522} & \textbf{0.484} & 0.102          & 0.093          & 0.053          & 0.049          \\
GPT-4.1           & 1.000 & 1.000 & 0.233          & 0.159          & 0.233          & 0.159          &  & 0.061          & 0.058          & 0.274          & 0.190          & 0.017          & 0.012          &  & 0.478          & 0.320          & 0.127          & 0.105          & 0.061          & 0.050          \\
Qwen3-8b-R        & 0.971 & 0.985 & 0.190          & 0.133          & 0.184          & 0.129          &  & 0.095          & 0.174          & 0.248          & 0.186          & 0.024          & 0.018          &  & 0.512          & 0.338          & 0.109          & 0.093          & 0.056          & 0.048          \\
Qwen3-32b-R       & 0.971 & 0.985 & 0.203          & 0.144          & 0.197          & 0.140          &  & \textbf{0.245} & \textbf{0.393} & 0.258          & 0.199          & \textbf{0.063} & \textbf{0.049} &  & 0.505          & 0.334          & 0.137          & \textbf{0.120} & \textbf{0.069} & \textbf{0.061} \\
Qwen3-8b          & 1.000 & 1.000 & 0.208          & 0.148          & 0.208          & 0.148          &  & 0.041          & 0.078          & \textbf{0.304} & \textbf{0.215} & 0.012          & 0.009          &  & 0.505          & 0.362          & 0.105          & 0.097          & 0.053          & 0.049          \\
Qwen3-32b         & 1.000 & 1.000 & 0.219          & 0.149          & 0.219          & 0.149          &  & 0.058          & 0.109          & 0.272          & 0.197          & 0.016          & 0.011          &  & 0.434          & 0.271          & 0.122          & 0.103          & 0.053          & 0.045          \\ \hline
\end{tabular}
}
\caption{Performance on AuditFraudBench. ``r1-c'': ROUGE1-conditional, ``rL-c'': ROUGEL-conditional, ``r1-g'': ROUGE1-gated, ``rL-g'': ROUGEL-gated. \label{tab:performancethreetasks}}
\end{table*}


\subsection{Result Analysis}

Table \ref{tab:performancethreetasks} reports model performance on the three tasks in AuditFraudBench. Overall, the results show that enforcement-grounded financial misinformation remains challenging for current LLMs. Although several models achieve high classification accuracy on Task 1, their ROUGE scores remain low, indicating that correct label prediction does not necessarily imply accurate explanation of the underlying accounting driver. This gap is more evident in Task 2 and Task 3, where both classification and explanation scores are limited. Since ROUGE-conditional is computed only on correctly classified examples and ROUGE-gated assigns zero explanation score to incorrectly classified examples, the gap between them reflects the end-to-end difficulty of jointly predicting the correct label and generating a faithful explanation. Across tasks, the results suggest that current LLMs still struggle to reason over financial figures, disclosure framing, restatement evidence, and enforcement-grounded misconduct mechanisms.

For \textbf{Task 1}, \textbf{Overview:} most models achieve near-perfect classification performance. Deepseek-v4-flash, GPT-5.5, GPT-4.1, Qwen3-8b, and Qwen3-32b all reach 1.000 accuracy and F1, while Deepseek-v4-pro, Qwen3-8b-R, and Qwen3-32b-R obtain 0.971 accuracy. This high performance should be interpreted carefully, because all Task 1 cases involve attribution errors in our enforcement-derived setting. Explanation metrics provide a stricter view: even the best model, GPT-5.5, reaches only 0.249 ROUGE-1 and 0.172 ROUGE-L, showing that models often fail to precisely identify the true accounting source of the performance change. \textbf{Effect of Model Scale and Capability.} In terms of model scale and capability, moving from Qwen3-8b to Qwen3-32b brings only marginal gains in explanation overlap, while the GPT and Deepseek series as a stronger model group do not consistently outperform the smaller Qwen models. For example, GPT-5.5 achieves the best ROUGE scores, but GPT-4.1 and Deepseek models are comparable to or weaker than Qwen3 variants on several explanation metrics. \textbf{Reasoning vs non-reasoning.} Reasoning variants also fail to bring consistent gains: Qwen3-8b-R and Qwen3-32b-R underperform their non-reasoning counterparts on ROUGE, and Deepseek-v4-pro performs worse than Deepseek-v4-flash. This suggests that Task 1 requires domain-specific accounting grounding rather than general reasoning or scale alone.

For \textbf{Task 2}, \textbf{Overview:} all models perform poorly, making Misleading Narrative Detection the most difficult task. Qwen3-32b-R achieves the best classification results, with 0.245 accuracy and 0.393 F1, and also obtains the highest ROUGE-gated scores. This indicates that reasoning-oriented models can help when the task requires identifying omitted context, selective emphasis, or favorable framing. However, the absolute scores remain low, showing that models still struggle to detect misleading impressions that are not explicitly false in isolation. \textbf{Effect of Model Scale and Capability.} Comparing model scale and capability, the GPT and Deepseek series do not show clear advantages over Qwen models. In fact, the best result comes from Qwen3-32b-R rather than the GPT or Deepseek series, and GPT-5.5, GPT-4.1, Deepseek-v4-pro, and Deepseek-v4-flash all obtain low classification scores. Within Qwen, scaling from 8b to 32b alone is not sufficient: Qwen3-32b does not improve over Qwen3-8b, while Qwen3-32b-R substantially improves over Qwen3-8b-R. \textbf{Reasoning vs non-reasoning.} The reasoning comparison is more favorable here than in Task 1. Qwen3-8b-R and Qwen3-32b-R outperform their non-reasoning counterparts in F1, especially Qwen3-32b-R, while Deepseek-v4-pro still does not improve over Deepseek-v4-flash. This suggests that reasoning helps narrative-level misinformation detection in some model families, but the benefit is not universal.

For \textbf{Task 3},  \textbf{Overview:} models show moderate but still limited ability to classify fraud patterns. GPT-5.5 performs best on classification, achieving 0.522 accuracy and 0.484 F1, while Qwen3-32b-R obtains the strongest explanation scores, including the best ROUGE-L-conditional and ROUGE-gated results. This divergence shows that correct fraud category prediction and faithful explanation generation are related but distinct capabilities. \textbf{Effect of Model Scale and Capability.} In terms of model scale and capability, the GPT and Deepseek series as a stronger model group perform competitively but still do not uniformly dominate. GPT-5.5 achieves the best classification results, but GPT-4.1 and Deepseek models are comparable to or weaker than Qwen models on several metrics. Within Qwen, scaling from 8b to 32b does not consistently help: Qwen3-32b performs worse than Qwen3-8b in accuracy and F1, while Qwen3-32b-R only slightly improves over Qwen3-8b-R in accuracy and slightly decreases in F1. \textbf{Reasoning vs non-reasoning.} Reasoning variants show mixed effects. Qwen3-32b-R substantially improves over Qwen3-32b, but Qwen3-8b-R does not clearly outperform Qwen3-8b; similarly, Deepseek-v4-pro improves over Deepseek-v4-flash in accuracy but has lower F1, suggesting possible bias toward frequent categories. These results indicate that fraud pattern classification is difficult because misconduct descriptions often contain overlapping signals, such as revenue timing combined with narrative distortion or estimate manipulation used for earnings smoothing.

Taken together, the results lead to three observations. First, classification metrics alone can overstate model capability, especially when labels are skewed or enforcement-driven; explanation metrics are necessary to assess whether the model understands the underlying financial misconduct. Second, reasoning-tuned models are most beneficial for narrative misleadingness detection, but their benefits are inconsistent across tasks and model families. Third, the GPT and Deepseek series do not reliably outperform smaller Qwen models, and increasing model size from Qwen3-8b to Qwen3-32b also does not guarantee improvement, suggesting that AuditFraudBench requires specialized financial and accounting reasoning beyond general model scale.

Additionally, we further provide a detailed error analysis of GPT-5.5 and DeepSeek-v4-flash in Appendix \ref{app:erroranalysis}.

\section{Conclusion}

In this work, we introduce AuditFraudBench, an enforcement-grounded benchmark for evaluating LLMs' ability to detect fraudulent financial misinformation in audited corporate reporting. Built from authentic filings, restated financial statements, MD\&A disclosures, and SEC AAERs, the benchmark evaluates three audit-relevant capabilities: identifying the true source of reported performance, detecting misleading disclosure narratives, and classifying misconduct mechanisms into known fraud patterns. Experiments on mainstream LLMs show that current LLMs still struggle to jointly reason over financial figures, management narratives, restatement evidence, and fraud mechanisms; larger or reasoning-oriented models do not consistently yield better performance. These findings highlight the need for more audit-aware, evidence-grounded LLM evaluation and development in high-stakes financial reporting settings.

\section*{Limitations}

This work has several limitations. 
First, due to resource and budget constraints, our evaluation covers open-source models up to 32B parameters, together with selected GPT and DeepSeek models. Therefore, the results may not fully reflect the performance of larger open-source models, newer proprietary systems. 
Second, although AuditFraudBench is constructed from authentic filings and enforcement materials, the dataset remains relatively small because enforcement-grounded fraud cases are scarce and require substantial manual verification. Expanding the benchmark to include more companies, fiscal periods, industries, and fraud types would improve coverage and statistical robustness. 
Third, the benchmark is primarily based on SEC filings and AAERs, which may limit its generalizability to other jurisdictions, regulatory regimes, and reporting standards. Fourth, our tasks focus on text and structured financial evidence extracted from filings, but do not fully capture the interactive and iterative nature of real audit procedures, where auditors request additional evidence, consult working papers, and revise judgments over time. 
Finally, our explanation evaluation relies partly on lexical overlap metrics such as ROUGE, which may not fully capture the correctness, faithfulness, or usefulness of accounting explanations. Future work could incorporate expert evaluation, semantic similarity metrics, and more fine-grained evidence attribution to better assess explanation quality.

\section*{Ethical Considerations}

AuditFraudBench is constructed from publicly available corporate filings and SEC enforcement materials, and is intended solely for research on evidence-grounded financial misinformation detection. Nevertheless, the benchmark concerns real companies and misconduct cases, so care should be taken to avoid using model outputs as definitive legal, investment, or audit judgments. LLM predictions may be incomplete or incorrect, and should not replace professional auditors, regulators, or financial experts. We also emphasize that the benchmark is designed to support responsible evaluation of LLMs, rather than to provide guidance for concealing, imitating, or optimizing fraudulent reporting behavior.



\bibliography{acl_latex}

\appendix

\section{Related Work \label{sec:relatedwork}}

\begin{table*}[t]
\centering
\small
\resizebox{\textwidth}{!}{
\begin{tabular}{llcccccc}
\toprule
\textbf{Category} & \textbf{Benchmark} & \textbf{Financial Filing} & \textbf{Narrative Disclosure} & \textbf{Accounting Rule} & \textbf{Fraud Pattern} & \textbf{Enforcement Evidence} & \textbf{Manipulation Type} \\
\midrule
Audit Benchmark
& FinAuditing \cite{wang2025finauditing} & \cmark & \xmark & \cmark & \xmark & \xmark & XBRL inconsistency \\
& Automated Audit \cite{wang2025automating} & \pmark & \xmark & \cmark & \xmark & \xmark & Statement error \\
& FinRule-Bench \cite{vignesh2026finrule} & \cmark & \xmark & \cmark & \xmark & \xmark & Rule violation \\
& FinMaster \cite{jiang2025finmaster} & \xmark & \xmark & \pmark & \xmark & \xmark & General workflow error \\
\midrule
Financial Misinformation
& Fin-Fact \cite{rangapur2025fin} & \xmark & \xmark & \xmark & \xmark & \xmark & Claim-level falsehood \\
& FinDVer \cite{zhao2024findver} & \cmark & \pmark & \xmark & \xmark & \xmark & Document-grounded inconsistency \\
& FISCAL \cite{sharma2025fiscal} & \xmark & \xmark & \xmark & \xmark & \xmark & Numerical claim inconsistency \\
& MFMD-Scen \cite{liu2026same} & \xmark & \xmark & \xmark & \xmark & \xmark & Scenario-sensitive misinformation \\
& RFCBench \cite{jiang2026all} & \xmark & \xmark & \xmark & \xmark & \xmark & Directional, numerical, sentiment, causal distortion \\
\midrule
\textbf{Ours}
& \textbf{AuditFraudBench} & \cmark & \cmark & \cmark & \cmark & \cmark & \textbf{Fraudulent disclosure manipulation} \\
\bottomrule
\end{tabular}
}
\caption{Comparison of related benchmarks. \cmark indicates full coverage, \pmark indicates partial coverage, and \xmark indicates absence. AuditFraudBench is the only benchmark that jointly connects audited filings, narrative disclosures, accounting mechanisms, and enforcement-grounded fraud evidence.}
\label{tab:related_benchmark_comparison}
\end{table*}

\subsection{Audit Benchmark}

Recent audit benchmarks have begun to assess LLMs’ ability to reason over structured financial information. FinAuditing \cite{wang2025finauditing} builds a taxonomy-aligned benchmark from real XBRL filings, testing semantic matching, relationship extraction, and mathematical reasoning over financial disclosures. Automating Financial Statement Audits \cite{wang2025automating} evaluates models through a five-stage audit framework combining real financial tables with synthesized transaction data, showing that LLMs can identify some statement errors but struggle to explain them, cite standards, or complete revisions. FinRule-Bench \cite{vignesh2026finrule} focuses on rule-based reasoning over real financial statements and human-curated accounting principles, evaluating rule verification, rule identification, and multi-violation diagnosis. FinMaster \cite{jiang2025finmaster} offers a broader benchmark across financial literacy, accounting, auditing, and consulting, revealing sharp performance degradation on complex multi-step workflows. While these benchmarks are useful for evaluating structured filing analysis and accounting-rule reasoning, they do not directly assess fraud-oriented disclosure analysis, such as detecting misleading narratives, distinguishing management’s stated explanations from the true drivers of performance, or identifying strategic distortions in corporate reporting.

\subsection{Financial Misinformation Detection}

Financial misinformation detection benchmarks mainly evaluate factual reliability and claim verification in financial contexts. Fin-Fact \cite{rangapur2025fin} introduces a multimodal fact-checking dataset of financial claims with truthfulness labels, ruling statements, and textual and visual evidence, showing that even advanced generative models struggle with multimodal financial verification. FinDVer \cite{zhao2024findver} focuses on explainable claim verification over long, hybrid financial documents, demonstrating that current LLMs remain far below expert-level performance in complex document-grounded settings. FISCAL \cite{sharma2025fiscal} proposes a synthetic-data framework for financial fact-checking and trains a lightweight verifier for numerical financial claims, showing that domain-specific synthetic data can improve efficient and robust verification. MFMDScen \cite{liu2026same} evaluates behavioral biases in multilingual financial misinformation detection across complex economic scenarios, while RFCBench \cite{jiang2026all} studies paragraph-level misinformation detection in financial news under reference-free and comparison-based settings. However, these benchmarks are generally not grounded in audited financial filings or enforcement cases, and therefore do not capture the distinctive mechanisms of corporate disclosure manipulation, such as revenue timing, expense deferral, earnings smoothing, accounting estimate manipulation, or misleading MD\&A framing.

\section{Social Impact \label{app:socialimpact}}

AuditFraudBench may contribute positively by promoting more reliable LLM systems for financial disclosure review, audit assistance, investor protection, and regulatory analysis. By focusing on misleading narratives and enforcement-grounded fraud mechanisms, it encourages the development of models that are more sensitive to high-risk financial misinformation beyond surface-level factual errors. At the same time, deployment in real-world financial settings should be cautious: overreliance on imperfect models could create false confidence, unfairly flag companies, or mislead users about audit risk. Therefore, systems evaluated with AuditFraudBench should be used as decision-support tools with human oversight rather than as autonomous financial or legal authorities.

\section{Error Analysis \label{app:erroranalysis}}

\begin{figure*}[htb]
\centering
  \includegraphics[width=2\columnwidth]{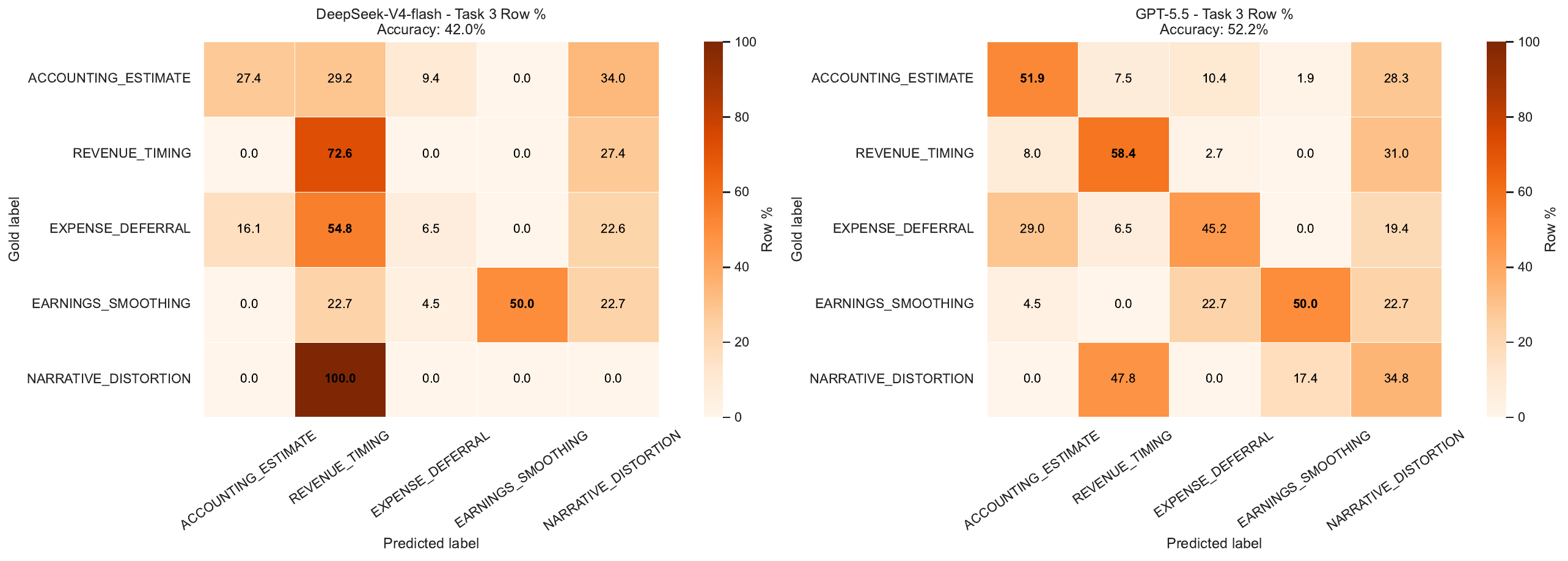}
  \caption{Confusion matrix on Task 3}
  \label{fig:Confusionmatrix}
\end{figure*}

\subsection{Confusion matrix on Task 3 (Figure \ref{fig:Confusionmatrix})}

The confusion matrices in Figure \ref{fig:Confusionmatrix} further reveal different error patterns between DeepSeek-V4-flash and GPT-5.5 on Task 3. DeepSeek-V4-flash achieves high recall for \textit{Revenue Timing}, correctly classifying 72.6\% of such cases, but it strongly over-predicts this category: all \textit{Narrative Distortion} cases and 54.8\% of \textit{Expense Deferral} cases are incorrectly assigned to \textit{Revenue Timing}. This suggests that DeepSeek-V4-flash tends to rely on dominant or salient fraud cues and has difficulty separating revenue-related manipulation from other misconduct mechanisms. In contrast, GPT-5.5 shows a more balanced pattern across categories, correctly identifying 51.9\% of \textit{Accounting Estimate}, 58.4\% of \textit{Revenue Timing}, 45.2\% of \textit{Expense Deferral}, and 50.0\% of \textit{Earnings Smoothing} cases. However, GPT-5.5 still struggles with \textit{Narrative Distortion}, correctly classifying only 34.8\% of cases and often confusing them with \textit{Revenue Timing} or \textit{Earnings Smoothing}. Overall, GPT-5.5 improves over DeepSeek-V4-flash by reducing category collapse and better distinguishing expense deferral and accounting estimate manipulation, but both models remain weak at recognizing narrative distortion as a distinct fraud pattern.

\subsection{Case studies of Deepseek-V4-flash}
\label{sec:c2-deepseek}

\subsubsection{DeepSeek Task 1: Profit Source Attribution}
\label{sec:err-deepseek-t1}

\textbf{Correct label, weak explanation: MQ Associates (AAER-2744, FY2002)}

\textbf{Prompt.}
Reported net income \$1.2\,M (period ending 2002-12-31); restated net
income $-$\$3.7\,M.
Management attribution: center count grew at a 7.6\% CAGR (1997--2000),
lagging the 12.3\% CAGR of FIC market revenue over the same period.

\textbf{Gold.}
Attribution is incorrect.
The 2002 net income swung more than 400\% from a \$1.2\,M profit to a
\$3.7\,M loss because the allowance for contractual adjustments on
accounts receivable was materially understated; MQ used a cash-trend
methodology instead of liquidation analysis until Q3~2004.

\textbf{Model.}
Attribution is incorrect: the management narrative discusses center growth
versus FIC market growth over 1997--2000, but the reported and restated
financials show a change in net income from \$1.2\,M to $-$\$3.7\,M for
the period ending 2002-12-31.
The narrative does not address any specific accounting errors, restatement
adjustments, or the reason for the large swing in net income.

\textbf{Analysis.}
The classification is correct.
The model rejects the passage because the growth discussion refers to
1997--2000 while the restatement concerns 2002, and because the figures move
from a \$1.2\,M profit to a \$3.7\,M loss without explanation.
Those observations are fair, but they stop short of the enforcement finding:
an understated allowance for contractual adjustments built with a cash-trend
method rather than liquidation analysis.
ROUGE-1 is 0.27, inflated by shared references to net income and restated
amounts rather than by overlap on the fraud mechanism.

\subsubsection{DeepSeek Task 2: Misleading Narrative Detection}
\label{sec:err-deepseek-t2}

\textbf{Correct classification, weak explanation: Premier Financial Bancorp, FY2003}

\textbf{Prompt.}
MD\&A passage for Premier Financial Bancorp: seven community bank
subsidiaries (\$24\,M--\$157\,M); as of December~31, 2003, Premier had
approximately \$62\,M in total assets, \$332\,M in total loans, and
\$455\,M in total deposits; allowance for loan losses maintained in
conformity with GAAP.

\textbf{Gold.}
Misleading.
The passage asserts GAAP compliance and systematic controls, but
management concealed loan losses, recorded fictitious loans, and
bump-dated loans to understate the required reserve, violating GAAP.

\textbf{Model.}
Misleading.
The passage states Premier had \$62\,M in total assets but \$332\,M in
loans and \$455\,M in deposits, which is mathematically inconsistent
because loans and deposits cannot exceed total assets by such magnitudes
under standard accounting principles.

\textbf{Analysis.}
The classification is correct.
The written rationale, however, rests on an internal inconsistency among
assets, loans, and deposits, not on the reserve-concealment scheme described
in the enforcement record (fictitious loans, bump-dating, and GAAP
violations).
Had the passage been numerically coherent, the model might still have missed
the misconduct.
ROUGE-1 is 0.28: terminology such as \emph{loans}, \emph{allowance}, and
\emph{GAAP} appears in both texts, but the causal accounts diverge.

\textbf{Correct classification, cross-model split: MQ Associates, AAER-2744, FY2002}

\textbf{Prompt.}
Passage from the MQ Associates 10-K: center growth CAGR (7.6\%,
1997--2000) versus FIC market revenue growth (12.3\%, 1997--2000);
followed by fragmented table-of-contents entries and the opening of the
business description section.

\textbf{Gold.}
Misleading.
The original filings materially overstated accounts receivable and
revenue because the allowance for contractual adjustments was
understated; MQ filed an 8-K on March~30, 2005 stating financial
statements for FY2002, FY2003, and Q1--Q3 2004 could not be relied upon.

\textbf{Model.}
Misleading.
The passage presents a growth rate for the number of centers (7.6\%) that
is lower than the market revenue growth rate (12.3\%), but it does not
clarify whether the company's own revenue growth is also lagging.

\textbf{Analysis.}
The classification matches the gold label.
DeepSeek flags the contrast between center and market growth; GPT-5.5 labels
the same excerpt \emph{not misleading} (see the GPT-5.5 case study below) and
treats the concessionary tone as balanced disclosure.
Neither model cites the understated allowance that drives the enforcement
record.

\textbf{Wrong classification: Tribune Company, AAER-2434, FY2002}

\textbf{Prompt.}
MD\&A: consolidated operating revenues increased 2\%, or \$131\,M, in
2002, due to improvements in publishing, broadcasting and entertainment
and interactive revenues; followed by table-of-contents debris and Item~1
company description.

\textbf{Gold.}
Misleading.
The passage omits fraudulent inflation of circulation figures that
materially overstated revenues.

\textbf{Model.}
Not misleading.
The passage provides a straightforward, factual description of a 2\%
revenue increase attributed to specific business segments, along with
standard corporate background information.

\textbf{Analysis.}
This is a false negative.
The excerpt reads plausibly on its own: segment-level growth and boilerplate
history contain no plainly false sentence.
The enforcement label instead reflects a material omission---circulation
fraud that inflated revenue---which never appears in the passage supplied
to the model.
The model searches for deceptive assertions rather than missing context, so
it concludes that the disclosure is not misleading.

\subsubsection{DeepSeek Task 3: Fraud Pattern Classification}
\label{sec:err-deepseek-t3}

\textbf{Correct category, weak rationale: MQ Associates, AAER-2744, FY2002}

\textbf{Prompt.}
FY2002 case narrative: ``MQ Associates' restatement largely resulted from
an understatement of its allowance for contractual adjustments for
accounts receivable.''

\textbf{Gold.}
\texttt{ACCOUNTING\_ESTIMATE}.
Standards violated: GAAP (AR allowances and revenue recognition);
Exchange Act \S\S\,13(b)(2)(A), 13(b)(2)(B), 15(d); Rules 12b-20,
15d-1, 15d-13.

\textbf{Model.}
\texttt{ACCOUNTING\_ESTIMATE}.
The restatement was caused by an understated allowance for contractual
adjustments on accounts receivable, which involves manipulation of a
discretionary accounting estimate rather than revenue timing or narrative
distortion.

\textbf{Analysis.}
The predicted category matches the gold label because the narrative names
an allowance shortfall.
The explanation restates that fact and rules out other categories, but does
not cite the legal standards recorded in the benchmark.
ROUGE-1 is 0.17: the model explains what was manipulated, while the
reference explanation lists what was violated.

\textbf{Wrong category (Tribune, AAER-2434, FY2002-Q3).}

\textbf{Prompt.}
Case narrative: ``From at least January~2002 to March~2004, Tribune
newspapers Newsday and Hoy falsely inflated paid circulation figures
through fictitious sales, free or below-minimum-price distribution
counted as paid, and shifting returns.''

\textbf{Gold.}
\texttt{NARRATIVE\_DISTORTION}.
Standards violated: Exchange Act \S\S\,13(a), 13(b)(2)(A), 13(b)(2)(B);
Rules 12b-20, 13a-1, 13a-13; GAAP.

\textbf{Model.}
\texttt{REVENUE\_TIMING}.
The misconduct involved inflating paid circulation through fictitious
sales and improper recognition of sales activity, fitting revenue timing.

\textbf{Analysis.}
The error follows from the rubric phrase ``fictitious sales,'' which also
appears in the case text.
Paid circulation, however, is an operational metric reported to industry
auditors and advertisers, not a GAAP revenue line.
Inflating those counts is narrative distortion of a non-financial
performance indicator, not mistimed revenue recognition.
Figure~\ref{fig:Confusionmatrix} (left panel) shows the same tendency at
scale: every gold \textit{Narrative Distortion} row is assigned to
\textit{Revenue Timing}.
ROUGE-1 is 0.00 because the predicted and reference explanations share
almost no substantive vocabulary.

\subsubsection{Summary (DeepSeek-V4-flash)}
\label{sec:err-deepseek-summary}

Across tasks, DeepSeek frequently reaches the right label while giving
reasons tied to wording or arithmetic in the prompt rather than to the
enforcement mechanism in the gold record.
On misleading-narrative detection, false negatives occur when MD\&A is
technically defensible in isolation; some true positives depend on
implausible balance-sheet ratios instead of the documented reserve fraud.
On fraud-pattern classification, recall for \textit{Revenue Timing} is
high, but \textit{Narrative Distortion} and several other categories are
routed into revenue-related labels when the narrative echoes sales or
recognition language.

\subsection{Case studies of GPT-5.5}

\subsubsection{GPT 5.5 Task 1: Profit Source Attribution}

\textbf{Correct label, weak explanation: MQ Associates (AAER-2744, FY2002)}
\label{sec:err-gpt55-t1}

\textbf{Prompt.}
Reported net income \$1.2\,M (period ending 2002-12-31); restated net
income $-$\$3.7\,M.
Management attribution: center count grew at a 7.6\% CAGR (1997--2000),
lagging the 12.3\% CAGR of FIC market revenue over the same period.

\textbf{Gold.}
Attribution is incorrect.
The 2002 net income swung more than 400\% from a \$1.2\,M profit to a
\$3.7\,M loss because the allowance for contractual adjustments on
accounts receivable was materially understated; MQ used a cash-trend
methodology instead of liquidation analysis until Q3~2004.

\textbf{Model.}
Attribution is incorrect: the management narrative discusses center growth
versus FIC market growth over 1997--2000, which does not address the 2002
restatement or explain the profit-to-loss change.

\textbf{Analysis.}
The verdict (correct vs.\ incorrect) is right.
The rationale is superficial.
The model correctly observes that the management narrative belongs to a
different time window (1997--2000) than the period being restated (2002),
and uses that temporal mismatch as the sole basis for rejection.
It never names the underlying accounting scheme, which is an understated AR
allowance arising from an improper cash-trend methodology. This is the
core finding in the gold explanation.
The explanation is essentially ``the narrative is off-topic,'' rather
than ``the narrative is off-topic \emph{because} management obscured a
specific reserve-estimation error.''
Mechanistic knowledge of the fraud is absent, and ROUGE-1 (0.43) inflates
because both texts share surface terms (\emph{net income}, \emph{restated},
dollar figures) while diverging entirely on the fraud explanation.

\subsubsection{GPT 5.5 Task 2: Misleading Narrative Detection}
\label{sec:err-gpt55-t2}

\textbf{Correct classification, weak explanation
  :Premier Bank, FY2003}\\

\textbf{Prompt.}
MD\&A passage for Premier Financial Bancorp: seven community bank
subsidiaries (\$24\,M--\$157\,M); as of December 31, 2003, Premier had
approximately \$62\,M in total assets, \$332\,M in total loans, and
\$455\,M in total deposits; allowance for loan losses maintained in
conformity with GAAP.

\textbf{Gold.}
Misleading.
The passage asserts GAAP compliance and systematic controls, but
management concealed loan losses, recorded fictitious loans, and
bump-dated loans to understate the required reserve, violating GAAP.

\textbf{Model.}
Misleading.
The passage reports \$62\,M in total assets against \$332\,M in loans and
\$455\,M in deposits.
Because loans are a primary asset and deposits fund assets, total assets
being far below both figures is implausible and suggests the disclosure is
erroneous or misleading.

\textbf{Analysis.}
The verdict is correct, but explanation focuses on to an arithmetic anomaly in
the passage (assets $\ll$ loans and deposits) rather than the actual
fraud.
The gold explanation describes a reserve-concealment scheme (fictitious
loans, bump-dating, GAAP violation); the model's explanation is purely a
balance-sheet arithmetic check.
The model stumbles on a surface-level inconsistency and treats it
as the evidence of fraud.
This ``shortcut'' reasoning is brittle: had the passage been numerically
consistent, the model might have rated it as not misleading, even though
the underlying GAAP violations would remain.
ROUGE-1 is 0.31, which reflects partial overlap in accounting terminology
(\emph{loans}, \emph{allowance}, \emph{GAAP}) but low overlap in the
causal mechanism.

\textbf{Wrong classification: MQ Associates, AAER-2744, FY2002}

\textbf{Prompt.}
Passage from the MQ Associates 10-K: center growth CAGR (7.6\%,
1997--2000) versus FIC market revenue growth (12.3\%, 1997--2000);
followed by fragmented table-of-contents entries (Items~8, 10, 13) and
the opening of the business description section.

\textbf{Gold.}
Misleading.
The original filings materially overstated accounts receivable and
revenue because the allowance for contractual adjustments was
understated; MQ filed an 8-K on March~30, 2005 stating financial
statements for FY2002, FY2003, and Q1--Q3 2004 could not be relied upon.

\textbf{Model.}
Not misleading.
The passage is a fragmented excerpt including table-of-contents material.
It explicitly states that center growth lagged market growth, which is a
balanced disclosure; there is no clear false statement or deceptive
omission in the text provided.

\textbf{Analysis.}
This is a false negative driven by the model's reasoning strategy for
misleadingness.

\emph{Tone-based assessment.}
The passage contains a statement that center count growth (7.6\% CAGR)
lagged FIC market revenue growth (12.3\% CAGR).
The model reads this as a voluntary concession of underperformance and concludes that a filing which openly acknowledges a
shortcoming is unlikely to be deceptive.
Self-critical or balanced language becomes a heuristic for
non-misleadingness, so the model returns ``not misleading.''

\emph{Assertion-checking versus completeness-checking.}
The passage is misleading not because any explicit claim in it is false,
but because it \emph{omits} material information: the same filing
overstated accounts receivable and revenue due to an understated allowance
for contractual adjustments.
The model's evaluation strategy is to scan for false or deceptive
assertions; it never asks whether the disclosure is \emph{complete}.
Because every statement in the passage is technically defensible on its
own, the model finds no evidence of deception.

The model would return the same verdict on any technically-true excerpt
drawn from a filing whose financials were materially misstated elsewhere,
because its reasoning has no mechanism to ask: ``What material information
should this passage have disclosed but did not?''

\subsubsection{GPT 5.5 Task 3: Fraud Pattern Classification}
\label{sec:err-gpt55-t3}

\textbf{Correct category, weak rationale: MQ Associates, AAER-2744, FY2002}\\

\textbf{Prompt.}
FY2002 case narrative: ``MQ Associates' restatement largely resulted from
an understatement of its allowance for contractual adjustments for
accounts receivable.''

\textbf{Gold.}
\texttt{ACCOUNTING\_ESTIMATE}.
Standards violated: GAAP (AR allowances and revenue recognition);
Exchange Act \S\S\,13(b)(2)(A), 13(b)(2)(B), 15(d); Rules 12b-20,
15d-1, 15d-13.

\textbf{Model.}
\texttt{ACCOUNTING\_ESTIMATE}.
The restatement was driven by an understated allowance for contractual
adjustments on AR, which is a discretionary accounting estimate/reserve
judgment rather than a revenue-timing or expense-deferral issue.

\textbf{Analysis.}
Classification is correct via strong keyword alignment: ``allowance,''
``contractual adjustments,'' and ``accounts receivable'' all point
unambiguously to the \texttt{ACCOUNTING\_ESTIMATE} category.
Rationale is logically sound but thin.
The model correctly rules out \texttt{REVENUE\_TIMING} and
\texttt{EXPENSE\_DEFERRAL} by negation, but offers no positive
characterization beyond paraphrasing the input.
The gold explanation is almost entirely a list of violated legal
provisions which the model never
mentions.
ROUGE-1 is only 0.14 despite correct classification, because the gold
and predicted explanations share almost no vocabulary: the model explains
\emph{what} was manipulated; the gold records \emph{what was violated}.
This gap reflects a structural mismatch between the model's output
(accounting rationale) and the benchmark's ground truth (legal citation).

\textbf{Wrong category (Tribune, AAER-2434, FY2002-Q3).}

\textbf{Prompt.}
Case narrative: ``From at least January~2002 to March~2004, Tribune
newspapers Newsday and Hoy falsely inflated paid circulation figures
through fictitious sales, free or below-minimum-price distribution
counted as paid, and shifting returns.''

\textbf{Gold.}
\texttt{NARRATIVE\_DISTORTION}.
Standards violated: Exchange Act \S\S\,13(a), 13(b)(2)(A), 13(b)(2)(B);
Rules 12b-20, 13a-1, 13a-13; GAAP.

\textbf{Model.}
\texttt{REVENUE\_TIMING}.
The misconduct centred on inflating paid circulation through fictitious
sales, counting free distributions as paid, and shifting returns, which
most closely aligns with fictitious sales and improper recognition of
sales activity.

\textbf{Analysis.}
The misclassification is caused by the phrase \emph{fictitious sales}
in the case narrative, which appears verbatim in the \texttt{REVENUE\_TIMING}
category definition (``fictitious sales, bill-and-hold transactions, channel
stuffing'').
The model anchors on this surface match and classifies accordingly.

However, Tribune's paid-circulation figures are an \emph{operational
disclosure metric}, not a GAAP revenue transaction.
Newspapers report circulation counts to industry bodies (ABC, BPA) and
use them as the basis for advertising rate cards; inflating those counts
is a form of false representation to third parties about a non-financial
key performance indicator.
The underlying violation is \texttt{NARRATIVE\_DISTORTION}: management
made false public statements about a business metric that investors and
advertisers rely on, without a corresponding misstatement of a recognised
revenue line.
There is no improper \emph{timing} of revenue recognition because
circulation is not itself a revenue-recognition event.

The error reveals a \emph{category-boundary ambiguity} in the
benchmark: when the act of falsification involves creating fictitious
transactions (fake circulation ``sales'') that inflate a reported
figure, the fraud superficially resembles \texttt{REVENUE\_TIMING}.
The distinction of whether the falsified figure appears in the GAAP income
statement or in a narrative/operational disclosure requires domain
knowledge that GPT 5.5 may not has.
ROUGE-1 is 0.00 because the predicted and gold explanations share no
substantive vocabulary (model: \emph{fictitious sales, recognition of
sales activity}; gold: legal citation strings).

\subsubsection{Summary (GPT-5.5)}
\label{sec:err-gpt55-summary}

Three consistent failure patterns emerge across all tasks.

\textbf{Surface reasoning over mechanism.}
In Task~1 and Task~3, the model correctly identifies the label by matching
surface keywords (wrong time period, ``allowance,'' ``fictitious sales'')
but never articulates the underlying accounting scheme.
Explanations are thin paraphrases of the input rather than causal
accounts of the fraud.

\textbf{Omission blindness in Task~2.}
The dominant error mode is false negatives: the model evaluates whether
the passage \emph{contains} a false or deceptive claim, but most Task~2
cases are misleading because material facts are \emph{omitted} from the
broader filing.
The model further treats candid or self-critical language as a positive
signal for honesty, reinforcing false negatives when the passage is
technically accurate but contextually incomplete.
Its reasoning strategy is assertion-checking, not completeness-checking,
and therefore systematically misses omission-based fraud.

\textbf{Category-boundary confusion in Task~3.}
Misclassifications concentrate at the boundaries between
\texttt{NARRATIVE\_DISTORTION} and the transaction-based categories
(\texttt{REVENUE\_TIMING}, \texttt{EARNINGS\_SMOOTHING}).
When the fraud involves falsifying a figure that \emph{drives} revenue
but is not itself a GAAP line item (e.g., circulation counts), the model
defaults to the transaction-level category.
Distinguishing the type of disclosure being distorted (GAAP income
statement vs.\ narrative/operational metric) requires contextual knowledge
that keyword-level reasoning does not supply.

\section{Prompt Templates for each task \label{sec:prompttemplates}}

\begin{figure}[htb]
\centering
\footnotesize
\fcolorbox{black}{gray!10}{
\begin{minipage}{0.45\textwidth}
\footnotesize

\textbf{Prompt for Task 1} \\ 

"""You are evaluating whether a company's stated explanation correctly attributes the source of a profit/restatement change.

Use only the information provided below.

Reported financials from the earliest disclosure: \\
\{compact\_json(reported\_financials)\}

Restated financials:\\
\{compact\_json(restated\_financials)\}

Management attribution narrative:\\
\{management\_attribution\}

Task:\\
Determine whether the management attribution narrative correctly explains the source of the change from the reported financials to the restated financials. If the attribution is incorrect or incomplete in a materially misleading way, explain why.

Return your answer as valid JSON with exactly these keys: \\
\{\{
  "attribution\_correct": true or false,\\
  "explanation": "brief explanation grounded in the reported financials, restated \\financials, and attribution narrative"
\}\}
"""

\end{minipage}
}
\caption*{Template 1: Prompt for Task 1}
\label{fig:propmt_task1}
\end{figure}

\begin{figure}[htb]
\centering
\footnotesize
\fcolorbox{black}{gray!10}{
\begin{minipage}{0.45\textwidth}
\footnotesize

\textbf{Prompt for Task 2} \\

"""You are a financial disclosure analyst.

Input passage: \\
{passage}

Task:\\
Determine whether this passage is a misleading/fraudulent narrative statement in the context of financial reporting, and explain your judgment.

Return your answer as valid JSON with exactly these keys:\\
\{\{
  "is\_misleading": true or false,
  "explanation": "brief explanation"
\}\}
"""

\end{minipage}
}
\caption*{Template 2: Prompt for Task 2}
\label{fig:propmt_task2}
\end{figure}

\begin{figure}[htb]
\centering
\footnotesize
\fcolorbox{black}{gray!10}{
\begin{minipage}{0.45\textwidth}
\footnotesize

\textbf{Prompt for Task 3} \\
\\

"""You are a forensic accounting analyst.

Case narrative: \\
{input\_text}

Classify the case into exactly one dominant fraud pattern category, using the following definitions:

(1) "REVENUE\_TIMING": "Assigned when the misconduct primarily involves improper timing or recognition of revenue, such as premature revenue recognition, fictitious sales, bill-and-hold transactions, channel stuffing, or recognizing revenue before required criteria were met.", \\
(2) "ACCOUNTING\_ESTIMATE": "Assigned when the misconduct primarily involves manipulation of accounting estimates or judgments, such as improper reserve releases, allowance manipulation, depreciation or amortization estimate changes, impairment judgments, or other discretionary estimates used to affect reported earnings.", \\
(3) "EARNINGS\_SMOOTHING": "Assigned when the misconduct primarily involves shifting income or expenses across periods to reduce earnings volatility or meet performance targets, including cookie-jar reserves or other practices designed to smooth reported results.", \\
(4) "EXPENSE\_DEFERRAL": "Assigned when the misconduct primarily involves delaying, capitalizing, or otherwise understating expenses or losses, such as improper capitalization of costs, failure to record liabilities, delayed impairment recognition, or deferral of ordinary operating expenses.", \\
(5) "NARRATIVE\_DISTORTION": "Assigned when the misconduct primarily involves misleading disclosures, omission of material information, or deceptive narrative framing intended to distort investors' understanding of the firm's performance, risks, or financial condition." \\

Task: \\
Choose the single best category and provide a brief explanation for your choice. The explanation is required.

Return your answer as valid JSON with exactly these keys: \\
\{\{
  "fraud\_category": "REVENUE\_TIMING" | "ACCOUNTING\_ESTIMATE" | "EARNINGS\_SMOOTHING" | "EXPENSE\_DEFERRAL" | "NARRATIVE\_DISTORTION",
  "explanation": "brief rationale for the selected category"
\}\}
"""

\end{minipage}
}
\caption*{Template 3: Prompt for Task 3}
\label{fig:propmt_task3}
\end{figure}

\end{document}